\documentclass[]{JHEP3}
\usepackage{epsfig}
\def\be{\begin{equation}}
\def\ee{\end{equation}}
\def\bea{\begin{array}}
\def\eea{\end{array}}
\def\beqa{\begin{eqnarray}}
\def\eeqa{\end{eqnarray}}
\def\beqas{\begin{eqnarray*}}
\def\eeqas{\end{eqnarray*}}

\def\bp{\begin{picture}}
\def\ep{\end{picture}}
\def\bc{\begin{center}}
\def\ec{\end{center}}
\def\bfig{\begin{figure}}
\def\efig{\end{figure}}

\def\bit{\begin{itemize}}
\def\eit{\end{itemize}}
\def\nn{\nonumber}
\def\f{\frac}

\def\[{\left[}
\def\]{\right]}
\def\({\left(}
\def\){\right)}

\def\..{\left.}
\def\.{\right.}
\def\tl{\tilde}
\def\ra{\rightarrow}
\def\la{\leftarrow}

\def\tm{\times}

\def\da{\dagger}

\def\la{\lambda}

\def\ep{\epsilon}

\def\pa{\partial}
\def\pr{\prime}

\title{Deflected Anomaly Mediated SUSY Breaking Scenario With General Messenger-Matter Interactions}
\author{Fei Wang$^{1,2}$\\
$^1$ School of Physics, Zhengzhou University, 450000,ZhengZhou
P.R.China\\
$^2$ State Key Laboratory of Theoretical Physics, Institute of Theoretical Physics,
                Chinese Academy of Sciences, Beijing 100080, P. R. China
}

\abstract{We propose to introduce general messenger-matter interactions in the deflected anomaly mediated SUSY breaking scenario. The most general form for the resulting soft parameters are derived. New interference terms between the GMSB type and AMSB type contributions are the unique feature of this scenario.  Messenger-matter interactions involving sleptons can be used to solve the tachyonic slepton problem and naturally lead to positive slepton masses regardless of the sign of deflection parameter. Besides, due to the new contributions, large $|A_t|$ that will not trigger color-breaking stop VEV are also possible in this scenario, thus can easily give the 125 GeV higgs which was discovered by LHC.  This type of deflected AMSB scenario need very few messenger species, thus can avoid possible non-perturbative gauge couplings below the GUT scale ( or Landau pole below the Planck scale ).

}

\begin{document}
\maketitle \indent
\newpage
\section{Introduction}

Low energy supersymmetry (SUSY) was regarded by many as one of the most appealing candidate for the TeV-scale new physics.
Not only the gauge hierarchy problem of Standard Model(SM) can be solved naturally with its SUSY extension,
 but also the dark matter puzzle can be explained elegantly. Besides, the gauge coupling unification which could not be exact
 in SM can however be successfully realized in the framework of low energy supersymmetry.
 Especially, the 125 GeV higgs boson which was discovered by both the ATLAS\cite{ATLAS:higgs} and CMS collaborations\cite{CMS:higgs} of the Large Hadron
 Collider(LHC), lies miraculously in the narrow $115-135$ GeV "window" predicted by the Minimal Supersymmetric Standard Model (MSSM).

Although SUSY is appealing, low energy SUSY seems to have some tension with LHC discoveries. In particular, no signals of SUSY particles had been observed at the the LHC so far besides the higgs boson. The LHC data had already set stringent constraints\cite{CMSSM1,CMSSM2} on certain CMSSM models:
$m_{\tilde g} > 1.5$ TeV for $m_{\tl{q}} \sim m_{\tl{g}}$, and $m_{\tl{g}}\gtrsim 1$ TeV for $m_{\tl{q}} \gg m_{\tl{g}}$. In fact, the experimental data reported by LHC agree quite well with the SM predictions.  No significant deviations from the SM had been observed in electroweak precision measurements as well as in flavor physics. So if low energy SUSY is indeed the new physics beyond the SM, its spectrum will display an intricate structure. So the origin of supersymmetry breaking is thus very crucial for phenomenology.

There are many ways to mediate the SUSY breaking effects from the hidden sector to the visible MSSM sector, the most well known ones are the gravity mediation\cite{SUGRA}, gauge mediation\cite{GMSB}, and anomaly mediation\cite{AMSB} mechanisms. The anomaly mediated SUSY breaking (AMSB) mechanism predicts a flavor-blind sparticle mass spectrum which is insensitive to any high energy theories\cite{deflect:RGE-invariance} and thus automatically solves the SUSY flavor problem. Unfortunately, the AMSB scenario, which is typically determined by $m_{3/2}$,  predicts tachyonic sleptons so that the minimal theory must be extended. There are several ways to tackle such tachyonic slepton problems\cite{tachyonslepton}. A very elegant solution is the deflected  AMSB\cite{deflect} scenario, in which messenger sectors are introduced in the AMSB to deflect the Renormalization Group Equation(RGE) trajectory and give new contributions to soft SUSY breaking terms. The tachyonic slepton problems can be solved with such deflection. On the other hand, relatively large messenger species are needed to give positive slepton masses in case of negative deflection parameter. However, more messenger fields may lead to non-perturbative gauge couplings below GUT scale (or Landau pole below Planck scale). Positively deflected AMSB\cite{okada}, which could do better, need special form of superpotential and is difficult for model building. So it is preferable to introduce less messenger fields to deflect the RGE trajectory that can give positive slepton masses.

  In this paper, we propose to introduce general messenger-matter interactions in deflected AMSB scenario. General messenger-matter interactions in GMSB had been studied in various papers\cite{gmsb-mm}. In our scenario, the slepton sector can receive additional contributions from both the messenger-matter interactions and ordinary deflected anomaly mediation to avoid tachyonic slepton masses. At the same time, additional contributions to trilinear coupling $A_t$ term which typically increase $|\tilde{A}_{t}|(\equiv A_t-\mu\cot\beta|)$ could be helpful to give 125 GeV higgs.

   This paper is organized as follows. We derive the forms of soft SUSY parameters for deflected AMSB scenario with general messenger-matter interactions in Sec 2. The SUSY spectrum for deflected AMSB with messenger-slepton-slepton interactions are given and studied in Sec 3. Sec 4 contains our conclusions.

\section{General Matter-Messenger Interactions in Deflected AMSB}
Ordinary AMSB is bothered with tachyonic slepton problems. An elegant way to solve such difficulty is the deflected anomaly mediation scenario which can change the RGE trajectory below the messenger thresholds. We can generalize the superpotential  in deflected AMSB scenario
 \beqa
 W=X\bar{\psi}_i\psi_i+W(X)
 \eeqa
 to a form that including general messenger-matter interactions
 \beqa
 W=\la_{\phi ij}X Q_iQ_j+ y_{ijk}Q_i Q_j Q_k+W(X)~.
 \eeqa
with the Kahler potential
\beqa
K=Z_U(\f{\mu}{\sqrt{\phi^\da\phi}}) Q_i^\da Q_i~.
\eeqa
The indices $'i,j'$ run over all MSSM and messenger fields and the subscripts $'U,D'$ denotes the case upon and below the messenger threshold, respectively.

 After integrating out the messenger fields, we have the general form for MSSM fields only
\beqa
{\cal L}=\int d^4\theta  Q_a^\da Z_D^{ab}(\f{\mu}{\sqrt{\phi^\da\phi}},\sqrt{\f{X^\da X}{\phi^\da \phi}}) Q_b+\int d^2\theta y_{abc}Q^aQ^bQ^c~,
\eeqa
 which can give additional contributions to soft supersymmetry breaking parameters. Here $'\phi'$ denotes the compensator field with Weyl weight 1.
  The analytic continuing threshold superfield $'X'$ will receive F-term from anomaly mediation and has the form $<X>=M+\theta^2 F_X$.

We can also expand the wave-function renormalization in power of $\theta$ with
\beqa
\tl{X}\equiv \f{X}{\phi}&=& \f{M+F_X\theta^2}{1+F_\phi\theta^2}\\
&=& M[1+(d+1)F_\phi\theta^2](1-F_\phi \theta^2)~,\\
&=&M(1+d F_\phi \theta^2)
\eeqa
The value of the deflection parameter $'d'$ is determined by the form of superpotential $W(X)$.
It is interesting to compare this result with gauge mediation scenario in which the dependence of Kahler metric on $X$ is given by
$<X>=M+F\theta^2$. The key difference between the gauge mediation contributions in deflected AMSB and ordinary GMSB is the appearance of the compensator field $X/\phi$.

We can naively obtain the soft SUSY breaking parameters by the replacement
\beqa
F_{\tl{X}}\ra M d F_\phi\ra \tl{X} d F_\phi~,
\eeqa
to transplant certain results from GMSB to our deflected AMSB scenario in addition to possible AMSB contributions. We will see soon that the interference between the (analog) gauge mediation type and pure anomaly mediation type terms also gives very important contributions.

After we expand the $Z^i$ to $\bar{\theta}^2\theta^2$ order,  we can canonically normalize the wavefunction with
\beqa
{Q}^\pr_i&\equiv& (Z^i_{D})^{1/2}\left(1-\f{F_\phi}{2}\theta^2\f{\pa}{\pa\ln\mu}\ln Z^i_D  +\f{F_{\tl{X}}}{2}\theta^2\f{\pa}{\pa \tl{X}}\ln Z_D^i\right) Q_i,~\nn\\
    &=& (Z^i_{D})^{1/2}\left[1-\f{F_\phi}{2}\theta^2\f{\pa}{\pa\ln\mu}\ln Z_D^i  +\f{d F_{\phi}}{2}\theta^2\f{\pa}{\pa \ln \tl{X}}\ln Z_D^i\right]Q_i,~
\eeqa
and arrive at
\beqa
Z^i_D Q_i^\da Q_i &=& \left[1+\bar{\theta}^2\theta^2\(\f{|F_\phi|^2}{4}\f{\pa^2 }{\pa (\ln\mu)^2}+\f{|F_{\tl{X}}|^2}{4}\f{\pa^2}{\pa |\tl{X}|^2}-\f{|F_\phi||F_{\tl{X}}|}{2}\f{\pa^2}{\pa\ln\mu\pa|\tl{X}|}\)\ln Z^i_D(\mu,|\tl{X}|)\right] Q_i^{\pr\da} Q_i^\pr~,\nn
\eeqa

So the leading order contributions to trilinear terms and scalar terms are
\beqa
\label{AMSB-formula}
\f{A_{abc}}{y_{abc}}&=&\sum\limits_{i=a,b,c}\left(-\f{1}{2}F_\phi\f{\pa}{\pa\ln \mu} +\f{d F_{\phi}}{2}\f{\pa }{\pa \ln |\tl{X}|}\right)\ln Z^{ii}_D({\mu},|\tl{X}|)~, \\
m^2_{ab}&=&\left(-\f{|F_\phi|^2}{4}\f{\pa^2 }{\pa (\ln\mu)^2}-\f{|F_{\tl{X}}|^2}{4}\f{\pa^2}{\pa |\tl{X}|^2}+\f{|F_\phi||F_{\tl{X}}|}{2}\f{\pa^2}{\pa\ln\mu\pa|\tl{X}|}\right)\ln Z_D^{ab}(\mu,|\tl{X}|)~,\nn\\
&=&\left[-\f{|F_\phi|^2}{4}\(\f{\pa \gamma^a }{\pa g_i}\beta(g_i)+\f{\pa \gamma^a}{\pa y_i}\beta(y_i)\)\right]+\left[-\f{d^2|F_{\phi}|^2}{4}\f{\pa^2}{\pa \ln|\tl{X}|^2}+\f{d|F_\phi|^2}{2}\f{\pa^2}{\pa\ln\mu\pa\ln|\tl{X}|}\right]\ln Z_D^{ab}(\mu,|\tl{X}|)~,\nn
\eeqa
Note that the last term is the unique feature of this deflected AMSB scenario which involve the interference between the pure anomaly and gauge mediation type contributions. Obviously, the soft scalar masses are not naively the sum of gauge mediation type and pure anomaly mediation type contributions.

The derivative of wavefunction $Z_{D}$ can be obtained from its integral expression
\beqa
Z_D(\ln \mu,M)=\int\limits^{\ln \mu}_{\ln X}d t^\pr \f{d Z_D(t^\pr,\ln X)}{dt^\pr}+\int\limits^{\ln X}_{\ln \Lambda_{UV}}d t^\pr \f{d Z_U(t^\pr)}{dt^\pr}
\eeqa
We follow the approaches in \cite{shih}  which uses the relation
\beqa
Z=V^\da V~,~~\f{d}{dt} V=-\gamma V~,~~~\Longrightarrow \f{d}{dt}Z=-2 V^\da \gamma V~,
\eeqa
and the expression
\beqa
\label{anomalous}
G_{ij}[Z(\ln\mu);\la(\ln\mu);g(\ln\mu)]&\equiv& -2V^\da \gamma V~,\nn\\
&=&-\f{1}{8\pi^2}\left(\f{1}{2}d_i^{kl}\la_{ikl}^* Z_{km}^{-1~*}Z_{ln}^{-1~*}\la_{jmn}-2c_r^i Z_{ij}g_r^2\right)~,
\eeqa
 to obtain\cite{chacko} at the messenger scale $\mu=|X|$
\begin{eqnarray}
\label{ddZ}
\f{\pa Z_D^{ab}(\ln\mu,|X|)}{\pa X}&=&\f{1}{2X}\Delta G^{ab}~,\\
\f{\pa^2 Z_D^{ab}(\ln\mu,|X|)}{\pa^2 |X|}&=&\f{1}{4|X|^2}\(\Delta\( \f{\pa G_{ab}}{\pa Z_{ij}}\)G_{ij}^{U}-\f{\pa G_{ab}^D}{\pa Z_{ij}^D}\Delta G_{ij}+\Delta\(\f{\pa G_{ab}}{\pa g_r}\)\beta_{g_r}^U-\f{\pa G_{ab}^D}{\pa g_r^D}\Delta \beta_{g_r}\)~.\nn
\end{eqnarray}
with $'\Delta(\cdots )'$ denoting the discontinuity of its followed expression. $d_i^{kl}$ is the standard multiplicity factor in the one-loop anomalous dimensions.

On the other hand, we must also calculate the interference terms between the anomaly mediation and gauge mediation.
From the integral expression, we have
\beqa
\f{\pa }{\pa \ln|\tl{X}|} Z_D^{ab}(\mu,|\tl{X}|)&=&\f{1}{2}\Delta G^{ab}+\int\limits^{\ln \mu}_{\ln X}d t^\pr \f{\pa}{\pa \ln|\tl{X}|}G_D^{ab}[Z_D(t^\pr,\tl{X});\la(t^\pr,\tl{X});g(t^\pr,\tl{X})]~.
\eeqa
So we can obtain that
\beqa
\f{\pa^2}{\pa\ln\mu\pa \ln |\tl{X}|} Z_D^{ab}(\mu,|\tl{X}|)&=&\f{\pa}{\pa \ln|\tl{X}|}G_D^{ab}[Z^D_{ij}(\ln\mu,\tl{X});\la(\ln\mu,\tl{X});g(\ln\mu,\tl{X})]~,\\
&=&\(\Delta(\beta_{\la})\f{\pa}{\pa \la}+\Delta(\beta_g)\f{\pa}{\pa g}+\sum\limits_{ij}\f{\pa Z^D_{ij}}{\pa\ln\tl{X}}\f{\pa}{\pa Z^D_{ij}}\)G_D^{ab}[Z_{ij}^D(\ln\mu);\la(\ln\mu);g(\ln\mu)]~. \nn
\eeqa
where
\beqa
\f{d g(\ln\mu,\ln X)}{d\ln|X|}=\Delta(\beta_{g})~,
\eeqa
with
\beqa
g(\ln\mu,\ln |X|)=g(\Lambda_{UV})+\int\limits_{\ln\Lambda_{UV}}^{\ln |X|} dt^\pr \beta_g^U(t^\pr)+\int\limits_{\ln|X|}^{\ln\mu} dt^\pr \beta_g^D(t^\pr,\ln |X|).
\eeqa
Similarly results hold for yukawa couplings. We can see that ordinary deflected AMSB is a specially case (with only $\Delta \beta_g$ non-vanishing ) of our general results.

 From the expression (\ref{anomalous}), we can obtain that
 \beqa
 &&\f{\pa Z^D_{ij}}{\pa\ln\tl{X}}\f{\pa}{\pa Z_{ij}^D}G_D^{ab}[Z_{ij}^D(\ln\mu);\la(\ln\mu);g(\ln\mu)]\nn\\
&&= \f{1}{8\pi^2}\[d_a^{kl}\la_{akm}^*\la_{bln}(Z^{-2}_{kl}Z^{-1}_{mn}\delta_i^k\delta_j^m\f{\Delta G^D_{kl}}{2}+Z^{-2}_{mn}Z^{-1}_{kl}\delta_i^m\delta_j^n\f{\Delta G^D_{mn}}{2})+2\delta_i^a\delta_j^b c_r^i g_r^2 \f{\Delta (G^{ij})}{2}\]~,~\nn\\
&&{\Delta (\beta_g)}\f{\pa}{\pa g_k}G_D^{ii}[Z_D^a(\ln\mu);\la(\ln\mu);g(\ln\mu)]=\f{1}{2\pi^2}  c_r^k \f{1}{16\pi^2} g_k^4  {\Delta} ({b_k})~,
 \eeqa

So we arrive at the final results for trilinear and scalar soft masses with general messenger sector at the messenger scale
\beqa
A_{a}=-\f{1}{2}G^D_{aa} F_\phi-\f{1}{32\pi^2} d_a^{ij}\Delta(|\la_{aij}|^2) d F_\phi~,
\eeqa
and the scalar soft masses
\beqa
\label{all}
m^2=m^2_{\rm AMSB}+m^2_{\rm gauge}+m^2_{\rm inter}~,
\eeqa
with
\beqa
(m^2)_{\rm AMSB}&=&\left[-\f{|F_\phi|^2}{4}\(\f{\pa \gamma^a }{\pa g_i}\beta(g_i)+\f{\pa \gamma^a}{\pa y_i}\beta(y_i)\)\right]~,\\
\label{interference}
(m^2_{ab})_{\rm inter}&=&\f{d F_\phi^2}{2}\left\{\f{}{}
-\f{1}{8\pi^2}\[d_a^{kl}\la_{akl}^*\la_{bmn}(\f{\Delta G^D_{km}}{2}+\f{\Delta G^D_{ln}}{2})+2 c_r^i g_r^2\f{\Delta G^D_{ab}}{2}\]\right.\nn\\
&+&\left.\f{1}{8\pi^2} 4 c_r^k \f{1}{16\pi^2} g_k^4  {\Delta} ({b_k})-G_D\f{\Delta G_D}{2}\right\}~,
\eeqa
and gauge mediation type contributions similar to \cite{shih}
\beqa
\label{GMSB}
(m^2_{ab})_{GMSB}&=&\f{d^2 F_\phi^2}{4}\f{1}{256\pi^4}\left[\f{1}{2}d_a^{ik}d_{i}^{lm}\(\Delta(\la_{aik}^*\la_{bjk})(\la_{ilm}\la_{jlm}^*)^U\)
-(\la_{aik}^*\la_{bjk})^D\Delta(\la_{ilm}\la_{jlm}^*)\nn\right.\\
&+&\left. \f{1}{4}d_{a}^{ij}d_b^{kl}\Delta(\la_{aij}^*\la_{cij})\Delta(\la_{ckl}^*\la_{bkl})-d_a^{ij}C_r^{aij}g_r^2\Delta(\la_{aij}^*\la_{bij})\right].
\eeqa
as well as the contributions from the last two term of (\ref{ddZ}) which are calculated to be
\beqa
\delta(m^2_{ab})_{GMSB}&=&\f{d^2F_\phi^2}{16}\f{\pa G_{ab}^-}{\pa g_r}\Delta \beta_{g_r}
=\f{d^2F_\phi^2}{128\pi^4}c_r^ig_i^4 \Delta b_i~.
\eeqa
We should note that the trilinear couplings for the third generation $A_t,A_b,A_\tau$ will in general receive new contributions at the messenger scale in addition to ordinary AMSB contributions.

 Additional soft terms will arise from the linear term of the expansion of the wavefunction
\beqa
\int d^4\theta Z_{ab}^D(\f{\mu}{\sqrt{\phi^\da\phi}};|\tl{X}|)\Phi_a^\da\Phi_b\supset \f{F_{\tl{X}}}{2}\f{\pa Z_{ab}^D}{\pa {\tl X}}F_{\Phi_a}^\da\Phi_b
-\f{F_{\phi}}{2}\f{\pa  Z_{ab}^D}{\pa {\ln\mu}} F_{\Phi_a}^\da\Phi_b+h.c.~,
\eeqa
which just compensates the change of variable from $'Z_D'$ to $'\ln Z_D'$ in (\ref{interference}).
 Integrating out the F-components of $\Phi_c$, the additional contributions to $ m^2_{ab}$ are
 \beqa
\delta m_{ab}^2&=&\f{|F_{\tl{X}}|^2}{4}\f{\pa Z_{ac}^D}{\pa {\tl{X}^*}}\f{\pa Z_{cb}^D}{\pa {\tl{X}}}
 +\f{|F_{\phi}|^2}{4}\f{\pa Z_{ac}^D}{\pa {\ln \mu}}\f{\pa Z_{cb}^D}{\pa {\ln\mu}}-
 \f{F^\da_{\tl{X}}F_{{\phi}}}{2}\f{\pa Z_{ac}^D}{\pa \tl{X}^*}\f{\pa Z_{cb}^D}{\pa  \ln\mu}
 -\f{F_{\tl{X}}F^\da_{{\phi}}}{2}\f{\pa Z_{ac}^D}{\pa \tl{X}}\f{\pa Z_{cb}^D}{\pa  \ln\mu}~,\nn\\
 &=&\f{|F_{\phi}|^2}{4}\left[d^2\f{\pa Z_{ac}^D}{\pa \ln{\tl{X}^*}}\f{\pa Z_{cb}^D}{\pa \ln{\tl{X}}}
 +\f{1}{4}\f{\pa Z_{ac}^D}{\pa {\ln \mu}}\f{\pa Z_{cb}^D}{\pa {\ln\mu}}-
 \f{d}{4}\f{\pa Z_{ac}^D}{\pa \ln\tl{X}^*}\f{\pa Z_{cb}^D}{\pa  \ln\mu}
 -\f{d}{4}\f{\pa Z_{ac}^D}{\pa \ln\tl{X}}\f{\pa Z_{cb}^D}{\pa  \ln\mu}~\right].
 \eeqa
 within which the first two term has already been included in $m^2_{\rm GMSB}$ and $m^2_{\rm AMSB}$, respectively. The remaining two terms are just the last term of the interference contributions in (\ref{interference}).
 \section{Positive Slepton Masses and Natural 125 GeV Higgs With Messenger-Slepton Interactions}
 Possible new messenger-matter interactions involving sleptons could give new contributions to slepton masses.
 We fit all the matter contents into $\bar{\bf 5}$ and ${\bf 10}$ representation of SU(5) GUT group. In order to deflect the RGE trajectory, we also introduce a messenger pair in $\Psi({\bf 5})$ and $\bar{\Psi}(\bar{\bf 5})$ representation of SU(5) with the following decomposition in terms of $SU(3)_c\tm SU(2)_L\tm U(1)_Y$ representation
 \beqa
 \Psi({\bf 5})&=&Q_\phi(1,2)_{1/2}\oplus U_\phi(3,1)_{-1/3}~,\nn\\
\bar{\Psi}(\bar{\bf 5})&=&\overline{Q}_\phi(1,\bar{2})_{-1/2}\oplus\overline{U}_\phi(\bar{3},1)_{1/3}~,
 \eeqa

 We introduce the following superpotential that involve messenger-MSSM-MSSM interaction (especially slepton-slepton-messenger interaction)
 \beqa
 W=X \overline{Q}_\phi Q_\phi + X \overline{U}_\phi U_\phi +\sum\limits_{i}\la_L^i {Q}_\phi L_{L,i} E_{L,i}^c+\sum\limits_{i}\la_Q^i {Q}_\phi Q_{L,i} D_{L,i}^c +W(X) ~,
 \eeqa
 with certain form of superpotential $W(X)$ for pseduo-moduli field $X$ to determine the deflection parameter $'d'$; superscript $'i'$ denotes the family indices.

From the previous general expressions for soft parameters, we can obtain the soft SUSY breaking parameters for slepton, stop masses and trilinear couplings at the messenger scale
\beqa
\f{{m}^2_{\tl{L}_L,i}}{F_\phi^2}&=&-\f{1}{256\pi^4}\(\f{99}{50}g_1^4+\f{3}{2}g_2^4\)+\f{d^2}{256\pi^4}\[4|\la_{L,i}|^4
-2|\la_{L,i}|^2\(\f{9}{10}g_1^2+\f{3}{2}g_2^2\)\],\nn\\
&+&\f{d^2}{128\pi^4}\(\f{3}{20}g_1^4+\f{3}{4}g_2^4\)+\f{d}{2}\left\{\f{1}{32\pi^4}\(\f{3}{20}g_1^4+\f{3}{4}g_2^4\)
-\f{1}{64\pi^4}\(\f{3}{20}g_1^2+\f{3}{4}g_2^2\)|\la_{L,i}|^2\right\}\nn\\
&+&\f{1}{64\pi^4}\(\f{3}{20}g_1^2+\f{3}{4}g_2^2\)^2 +\f{d}{128\pi^4}|\la_{L,i}|^2\(\f{3}{20}g_1^2+\f{3}{4}g_2^2\).\\
&\approx& \f{d^2}{64\pi^4}|\la_{L,i}|^4.\nn\\
\f{m^2_{\tl{E}_{L,i}^c}}{F_\phi^2}&=&-\f{1}{256\pi^4}\(\f{198}{25}g_1^4\)
+\f{d^2}{256\pi^4}\[8|\la_{L,i}|^4-4|\la_{L,i}|^2\(\f{3}{5}g_1^2\)\]+\f{d^2}{128\pi^4}\(\f{3}{5}g_1^4\),~\nn\\
&+&\f{d}{2}\left\{\f{1}{32\pi^4}\(\f{3}{5}g_1^4\)-\f{1}{32\pi^4}\(\f{3}{5}g_1^2\)|\la_{L,i}|^2\right\}+\f{1}{64\pi^4}\f{9}{25}g_1^4 +\f{d}{64\pi^4}|\la_{L,i}|^2\f{3}{5}g_1^2\nn\\
&\approx&  \f{d^2}{32\pi^4}|\la_{L,i}|^4.\\
\f{{m}^2_{\tl{Q}_L^3}}{F_\phi^2}&=&\f{1}{256\pi^4}\(8g_3^4-\f{11}{50}g_1^4-\f{3}{2}g_2^4\)
+\f{d^2}{256\pi^4}\[6|\la_{Q,3}|^4-2|\la_{Q,3}|^2\(\f{7}{30}g_1^2+\f{3}{2}g_2^2+\f{8}{3}g_3^2\)\],~\nn\\
&+&\f{d}{2}\left\{\f{1}{32\pi^4}\(\f{1}{60}g_1^4+\f{3}{4}g_2^4+\f{4}{3}g_3^4\)
-\f{1}{64\pi^4}\(y_t^2+y_b^2+\f{1}{60}g_1^2+\f{3}{4}g_2^2+\f{4}{3}g_3^2\)|\la_{Q,3}|^2\right\}\nn\\
&+&\f{d^2}{128\pi^4}\(\f{1}{60}g_1^4+\f{3}{4}g_2^4+\f{4}{3}g_3^4\)+\f{1}{64\pi^4}\(\f{1}{60}g_1^2+\f{3}{4}g_2^2+\f{4}{3}g_3^2-\f{y_t^2}{2}-\f{y_b^2}{2}\)^2\nn\\
&+&\f{d}{256\pi^4}\[-(y_t^2+y_b^2)+\f{1}{30}g_1^2+\f{3}{2}g_2^2+\f{8}{3}g_3^2\]|\la_{Q,3}|^2\\
&\approx&\f{g_3^4}{32\pi^4}+\f{d^2|\la_{Q,3}|^2}{128\pi^4}\(|\la_{Q,3}|^2-\f{8}{3}g_3^2\)
+\f{(d^2+2d)}{128\pi^4}\f{4g_3^4}{3}+\f{1}{64\pi^4}\(\f{4}{3}g_3^2-\f{y_t^2}{2}\)^2-\f{3d}{256\pi^4}y_t^2|\la_{Q,3}|^2~,\nn \\
\f{m^2_{\tl{b}_L^c}}{F_\phi^2}&=&\f{1}{256\pi^4}\(8g_3^4-\f{22}{25}g_1^4\)
+\f{d^2}{256\pi^4}\[12\la_{Q,3}^4-4\la_{Q,3}^2\(\f{7}{30}g_1^2+\f{3}{2}g_2^2+\f{8}{3}g_3^2\)\],~\nn\\
&+&\f{d^2}{128\pi^4}\(\f{1}{15}g_1^4+\f{4}{3}g_3^4\)+\f{d}{2}\left\{\f{1}{32\pi^4}\(\f{1}{15}g_1^4+\f{4}{3}g_3^4\)
-\f{|\la_{Q,3}|^2}{32\pi^4}\(2y_b^2+\f{1}{15}g_1^2+\f{4}{3}g_3^2\)\right\}\nn\\
&+&\f{1}{64\pi^4}\(\f{1}{15}g_1^2+\f{4}{3}g_3^2-y_b^2\)^2
-\f{d}{128\pi^4}|\la_{Q,3}|^2\(2y_b^2-\f{2}{15}g_1^2-\f{8}{3}g_3^2\)~,\nn\\
&\approx&\f{g_3^4}{32\pi^4}+\f{d^2}{64\pi^4}\(3\la_{Q,3}^4-\f{8}{3}g_3^2\la_{Q,3}^2\)+\f{(d^2+2d)}{96\pi^4}g_3^4+\f{1}{36\pi^4}g_3^4~. \\
\f{m^2_{\tl{t}_L^c}}{F_\phi^2}&=&\f{1}{256\pi^4}\(8g_3^4-\f{88}{25}g_1^4\)-\f{d^2}{128\pi^4}y_t^2|\la_{Q,3}|^2+\f{d}{64\pi^4}\(\f{4}{15}g_1^4+\f{4}{3}g_3^4\)~,~\nn\\
&+&\f{d^2}{128\pi^4}\(\f{4}{15}g_1^4+\f{4}{3}g_3^4\)+\f{1}{64\pi^4}\(\f{4}{15}g_1^2+\f{4}{3}g_3^2-y_t^2\)^2~,\nn\\
&\approx&\f{g_3^4}{32\pi^4}-\f{d^2}{128\pi^4}y_t^2|\la_{Q,3}|^2+\f{(d^2+2d)}{96\pi^4}g_3^4+\f{1}{64\pi^4}\(\f{4}{3}g_3^2-y_t^2\)^2,\\
\f{A_{L_{L,i}}}{F_\phi}&=&-\f{d}{16\pi^2}|\la_{L,i}|^2~,~
\f{ A_{E_{L,i}^{c}}}{F_\phi}=-\f{d}{8\pi^2}|\la_{L,i}|^2~,
\f{A_{Q_{L,3}}}{F_\phi}=-\f{1}{16\pi^2}\({y_t^2+y_b^2}-\f{8}{3}g_3^2\)-\f{d}{16\pi^2}|\la_{Q,3}|^2~,~\nn\\
~\f{A_{b^c_{L}}}{F_\phi}&=&-\f{1}{16\pi^2}\(2{y_b^2}-\f{8}{3}g_3^2\)-\f{d}{8\pi^2}|\la_{Q,3}|^2~,
\eeqa
 Other soft parameters at the messenger scale which do not receive new contributions from messenger-matter interaction are not shown explicitly here.

 In previous formulas, we neglect the contributions involving  yukawa couplings for lepton and quarks other than $y_t$ as well as gauge couplings other than $g_3$. We also neglect terms corresponding to $\la\Delta(\beta_\la)$ which can only contribute sub-leading terms involving $y_b^2\la^2$.  In our calculation, we use the following anomalous dimension of each superfields
\beqa
\bea{|c|c|c|}
\hline
\Phi &(C_3,C_2,C_1) &\f{1}{2}d_{i}^{kl}\la_{ikl} \la_{ikl}\\
\hline
L_L^i& (0,\f{3}{4},\f{3}{20})&y_\tau^2\\\hline
(E_L^c)_i&(0,0,\f{3}{5})&2y_\tau^2\\\hline
Q_L^i&(\f{4}{3},\f{3}{4},\f{1}{60})&y_t^2+y_b^2\\\hline
(D_L^c)_i&(\f{4}{3},0,\f{1}{15})&2y_b^2\\\hline
(U_L^c)_i&(\f{4}{3},0,\f{4}{15})&2y_t^2\\
\hline
\eea\eeqa
and the relevant quadratic Casimirs of the couplings
\beqa
\bea{|c|c|c|}
\hline
{\cal O}   & (d_\phi,d_2,d_3)  & (C_r^{1},C^2_r,C^3_r) \\
\hline
{Q}_\phi L_{L,i} E_{L,i}^c &(1,1,2) &(\f{9}{10},\f{3}{2},0)\\
\hline
{Q}_\phi Q_{L,i} D_{L,i}^c&(3,1,2)& (\f{7}{30},\f{3}{2},\f{8}{3})\\
\hline
\eea
\eeqa
with $C_r^i=C_r^i(\phi_1)+C_r^i(\phi_2)+C_r^i(\phi_3)$. The changes of beta functions after passing messenger threshold are given by $$ (\Delta b_g^3, \Delta b_g^2,\Delta b_g^1)=(1,1,1).$$

 From our previous expressions, we have the following discussions:
 \bit
 \item In ordinary AMSB, the notorious tachyonic slepton masses can be naturally solved in our scenario. In addition to ordinary deflected AMSB contributions involving gauge couplings, the soft slepton masses receive dominant contributions from matter-messenger couplings.
  We can see that the slepton soft masses are always positive (with relatively large messenger-matter coupling) for both negative and positive deflection parameter $d$.
     This is another advantage of our scenario because large messenger species have to be introduced to give positive slepton masses in deflected AMSB scenario with negative deflection parameter as well as for positive deflection parameter $d\lesssim 2$.  In our scenario, one messenger specie can be enough to make slepton masses positive.
 \item The value of trilinear coupling $A_t$ which is
 \beqa
 A_t=A_{Q_{L,3}}+A_{t^c_{L}}+A_{H_u}~,
 \eeqa
receives additional contribution involving the couplings $\la_{Q,3}$ from $A_{Q_{L,3}}$. Depending on the sign of deflection parameter, the $A_t$ terms can be either positive or negative. Relatively large $\la_{Q,3}$ will easily lead to large value of $|A_t|$ which can naturally give a large higgs masses that can be compatible with LHC discovery. This is obviously from the formula
\beqa
m_{h}^{2}\simeq m_{Z}^{2}\cos^{2}2\beta+\frac{3m_{t}^{4}}{4\pi^{2}v^{2}}
\left[\log\frac{M_{\mathrm{SUSY}}^{2}}{m_{t}^{2}}+\frac{\tilde{A}_{t}^{2}}{M_{\mathrm{SUSY}}^{2}}\left(1-\frac{\tilde{A}_{t}^{2}}{12M_{\mathrm{SUSY}}^{2}}\right)\right], \eeqa
with $M_{\mathrm{SUSY}}=\sqrt{m_{\tilde{t}_{1}}m_{\tilde{t}_{2}}}$ the geometric mean of stop masses. We can either choose $M_{\mathrm{SUSY}}/m_{t}\gg1$ which results in a relatively heavy SUSY spectrum or $M_{\mathrm{SUSY}}/m_{t}>1$ and $\tilde{A}_{t}/M_{\mathrm{SUSY}}>1$ with large stop mixing to increase the loop contributions to higgs mass. The stop masses must be larger than 10 TeV in case of no stop mixing. Obviously, large $\tl{A}_t$ is preferable for low energy SUSY.

 \item The soft stop masses can also be changed with general messenger-matter interactions. We can see from our expression that the $m^2_{\tl{t}_L}$ tends to receive positive contributions (although it is also possible for the new contribution to be negative) while $m^2_{\tl{t}_L^c}$ tends to receive negative contributions. So it is possible to choose proper deflection parameter so that the geometric average of stop masses decrease.
     This again can be helpful for our scenario to be compatible with the 125 GeV higgs. Besides, relatively large $A_t$ and light stop can reduce the fine-tuning of the theory. In fact, radiatively natural SUSY\cite{rnaturalsusy} type spectrum can be  realized in our scenario.

\item As noted previously, large messenger species are necessary ($N_m\geq 4$) to give positive slepton masses in ordinary deflected AMSB scenario with negative deflection parameter. To obtain large positive slepton masses need even more messenger species which could result (for $N_m\geq 6$) in non-perturbative gauge couplings below the GUT scale (or Laudau pole below the Planck scale). Similar condition holds for the case of small positive deflection parameters. In our scenario, even $N_m=1$ can lead to positive slepton masses disregard of the sign of deflection parameter $'d'$. So possible non-perturbative gauge couplings below GUT scale (or Landau pole below Planck scale ) will naturally be avoid in our scenario.

\item  We use SuSpect2.0\cite{suspect} to obtain the low energy spectrum of our scenario with messenger scale input ( here $M=1.0\times 10^{8}$ GeV). We show in Table 1,2 two benchmark points with one messenger specie $N_m=1$ for negative (and positive) deflection parameters, respectively. The tachyonic slepton problems which bother ordinary AMSB scenario are absent in our scenario for both sign of the deflection parameters with $N_m=1$. Besides, possible color-breaking stop VEV  related to large $A_t$ are checked to be absent. Successful EWSB, typical constraints from precision measurments and $b\ra s\gamma$ etc \cite{bsgamma,gmu-2}are checked to satisfy the experimental constraints. We can see that 125 GeV higgs can easily be obtained in our scenario with both positive and negative deflection parameters.
    \begin{table}[h]\caption{The messenger scale inputs and the low energy spectrum for benchmark point I with $d<0$. All the quantities with mass dimension are in GeV. }
\centering
\begin{tabular}{|c|c|c|c|c|}
\hline
$(\la_L^i,~\la_Q^3)$    & $(d,N_m)$         & M           & $F_\phi$           & $(tan\beta, sig(\mu))$ \\ \hline
$(1.0,~1.0)$       &  (-1.0,~1)      & $1.0\tm 10^8$    & $2.0\tm 10^5$    & (10.0~,~+)       \\ \hline
\hline
$m_{\tl{t}_{1}} $   & $ m_{\tl{t}_{2}} $ &      $m_{\tl{b}_{1}} $       &  $m_{\tl{b}_{2}} $          & $M_{\tl{g}}$             \\ \hline
3521             &      4120        &   4093           & 5350       &   2879        \\ \hline
$m_{\tilde{\tau}_1}$   & $m_{\tilde{\tau}_2}$ & $m_{\tilde{\nu}_\tau}$  & $m_{\tilde{e}_1}$     & $m_{\tilde{e}_2}$ \\ \hline
1875            &   2601      &  2599      & 2606             & 1890              \\ \hline
$m_h$                 & $m_H$              & $m_A$                 & $M_{\tl{\chi}_1^0}$               & $M_{\tl{\chi}_2^0}$                    \\ \hline
125.23            & 2762.2             & 2762.1            & 1162       &   2131               \\ \hline
\hline
 $M_{\tl{\chi}_3^0}$ & $M_{\tl{\chi}_4^0}$ &$Br(B\rightarrow X_S\gamma)$& $g_\mu-2$           & $\Delta \rho$           \\ \hline
3004 &  3007  &    $ 3.234\times 10^{-4}$  & $2.158\times 10^{-11}$            &     $0.221\times 10^{-3}$                  \\ \hline
\end{tabular}
\end{table}

 \begin{table}[h]\caption{The messenger scale inputs and the low energy spectrum for benchmark point II with $d>0$. All the quantities with mass dimension are in GeV. }
\centering
\begin{tabular}{|c|c|c|c|c|}
\hline
$(\la_L^i,~\la_Q^3)$    & $(d,N_m)$         & M           & $F_\phi$           & $(tan\beta, sig(\mu))$ \\ \hline
$(1.0,~1.0)$       &  ( 0.5,~1)      & $1.0\tm 10^8$    & $2.1\tm 10^5$    & (10.0~,~+)       \\ \hline
\hline
$m_{\tl{t}_{1}} $   & $ m_{\tl{t}_{2}} $ &      $m_{\tl{b}_{1}} $       &  $m_{\tl{b}_{2}} $          & $M_{\tl{g}}$             \\ \hline
4550            &       4901        &   4859           & 5831       &   5599        \\ \hline
$m_{\tilde{\tau}_1}$   & $m_{\tilde{\tau}_2}$ & $m_{\tilde{\nu}_\tau}$  & $m_{\tilde{e}_1}$     & $m_{\tilde{e}_2}$ \\ \hline
1962             &    2660       & 2659      & 2664             &  1972               \\ \hline
$m_h$                 & $m_H$              & $m_A$                 & $M_{\tl{\chi}_1^0}$               & $M_{\tl{\chi}_2^0}$                    \\ \hline
126.18            & 2939.8             & 2939.8          &333       &   1805                \\ \hline
\hline
 $M_{\tl{\chi}_3^0}$ & $M_{\tl{\chi}_4^0}$ &$Br(B\rightarrow X_S\gamma)$& $g_\mu-2$           & $\Delta \rho$           \\ \hline
2939 & 2940  & $3.248\times 10^{-4}$  & $ 2.08\times 10^{-11}$            &      $0.6622\times10^{-3}$                 \\ \hline
\end{tabular}
\end{table}

\eit

We can also introduce messenger-messenger-matter type interactions for sleptons. Again, due to additional positive contributions from messenger-matter couplings, positive slepton masses and 125 Gev higgs can naturally be obtained. Detailed discussions of such possibility will be given in our subsequent study.

\section{\label{sec-3}Conclusions}
 We propose to introduce general messenger-matter interactions in the deflected anomaly mediated SUSY breaking scenario. The most general form for the resulting soft parameters are derived. New interference terms between the GMSB-type and AMSB-type contributions are the unique feature of this scenario. Messenger-matter interactions involving sleptons can be used to solve the tachyonic slepton problem and naturally lead to positive slepton masses regardless of the sign of deflection parameter. Besides, due to the new contributions, large $|A_t|$ that will not trigger color-breaking stop VEV are also possible in this scenario, thus can easily give the 125 GeV higgs which was discovered by LHC. This type of deflected AMSB scenario need very few messenger species, thus can avoid possible non-perturbative gauge couplings below GUT scale ( or Landau pole below the Planck scale ).

\begin{acknowledgments}
 This work was supported by the
Natural Science Foundation of China under grant numbers 11105124, 11105125; by the Open Project Program of State Key Laboratory of
Theoretical Physics, Institute of Theoretical Physics, Chinese Academy of Sciences, P.R.
China (No.Y5KF121CJ1); by the Innovation Talent project of Henan Province under grant
number 15HASTIT017 and the Young-Talent Foundation of Zhengzhou University.
\end{acknowledgments}

\end{document}